# Selfie: User-defined Sensitive Memory Protection and Recovery

Pengfei Sun
Rutgers University
pengfei.sun@rutgers.edu

Saman Zonouz
Rutgers University
saman.zonouz@rutgers.edu

*Abstract*—Different users always have different requirement for sensitive memory definition. It is not flexible for aborting program execution once detecting memory corruption. Because the users may loose some sensitive data. We presented Selfie, a hybrid solution to provide one flexible solution to protect the sensitive memory according to users' requirements in runtime. Finally, Selfie can provide one solution to decide whether execution needs to be recovered. If the memory corruption doesn't belong sensitive memory, Selfie provides symbolic solver that can help figure out whether the memory corruption can affect the sensitive memory in future.

## I. Introduction

Many programs are still written using type-unsafe languages such as C and C++ for performance and compatibility reasons. However, these languages don't provide intrinsic guarantees about memory safety [1], [2], [3]. Lack of memory safety, in turn, forms the basis for attacks in the form of code injection [4] and code reuse [5]. Once memory errors happen, the interesting sensitive memory need to keep integrity and recovery from users view. Most time, memory corruption can damage the memory integrity. Memory corruption occurs when a program breaks type safety and writes to an unintended location, potentially corrupting the data. Because all memory locations are equally accessible to all store instructions in a program, current approaches to providing safety from memory corruption in C and C++ require that every store in a program be either statically or dynamically checked for correctness. That means it will incur source code or binary code modification. As a result, there are significant challenges to existing approaches that limit their practical application.

Assuming the source code is available, it is not easy for developers to define sensitive memory for users. Because different users always have different views for sensitive memory. Especially, if the developer makes too much protection which will increase the performance overhead. Meanwhile, most of time, it is impossible for users to get the source code of commercial software. So it is not practical to custom the sensitive memory protection. Even if there are good solution to define the sensitive memory for different users, the another challenge will be how to detect the memory corruptions and recognize whether the memory corruptions will affect the sensitive memory and whether the clean state before memory corruptions need to be recovered. There already are many solutions about how to detect memory corruptions. A number of approaches that attempt to provide type and memory safety for C and C++ programs are fail-stop, aborting program execution once detecting an error [6], [7], [8]. Valgrind [9] use binary rewriting or emulation to dynamically detect memory corruptions in unaltered programs. Diehard [10] is a runtime system that can provide probabilistic memory safety. Fail-stop means the user will loose all sensitive memory once memory corruptions. However, memory corruptions detection are not the only requirements for users. The importance is how to response and protect the sensitive memory when there are memory corruptions. Sensitive memory deletion is good to avoid sensitive data leaking, however, in some sense, it is not good for users when they don't have any copies for the sensitive data.

In this paper, we present an alternate approach to reducing the impact of sensitive heap memory corruptions on C and C++ programs without modifying source code information or rewriting binary. Because most of sensitive data is in the heap. So we focus on heap memory not stack. Our main contributions are as follows:

- Selfie provide one flexible sensitive memory protection and recovery solution without modifying source code and rewriting binary. Selfie provide one modified memory allocator for C and C++ programs, which can be loading in runtime according to users' requirements which memory should be sensitive.

- Selfie analyze the possible heap memory corruption vulnerabilities. Selfie balance the pin tool to provide a practical solution to detect the memory corruptions. The importance is to figure out the root cause of the memory corruptions. So Selfie provides data structure extracting and dynamic program slicing to figure out the root cause of the memory corruptions.

- Selfie provide one solution to decide whether execution needs to be recovered. If the memory corruption doesn't belong sensitive memory, Selfie provides symbolic solver that can help figure out whether the memory corruption can affect the sensitive memory in future. Finally, Selfie can do execution recovery according to the above analysis.

## II. Heap Memory Corruption

There are a lot of memory corruption vulnerabilities in software. Heap memory corruption is very serious vulnerability currently. Out there in the wild, many memory allocators are available, such as dlmalloc, jemalloc, tcmalloc, malloc(glibc). The whole of paper will base on glibc malloc heap memory allocator. It will be easy to expand to other memory allocators.

Next, we will do simple introduction about glibc malloc memory allocator. And then we talk about the current heap memory corruptions.

*A. Heap Memory Allocation*

Each memory allocation is internally represented by **chunk**. A **chunk** consists of metadata and the memory returned to the program. All these chunks are saved on the heap, which is a memory region capable of expanding when new memory is requested. Similarly, the heap can shrink once a certain amount of memory has been freed.

```
struct malloc_chunk {
  INTERNAL_SIZE_T     prev_size;
  INTERNAL_SIZE_T     size;

  struct malloc_chunk* fd;
  struct malloc_chunk* bk;

  struct malloc_chunk* fd_nextsize;
  struct malloc_chunk* bk_nextsize;
};
```

For allocated chunk in heap, the memory layout of the heap is as follows:

- **prev_size**: If the previous chunk is free, this field contains the size of previous chunk. Else if previous chunk is allocated, this field contains previous chunk's user data.
- **size**: This field contains the size of this allocated chunk. Last 3 bits of this field contains flag information.
  - **PREV_INUSE(P)**: This bit is set when previous chunk is allocated.
  - **IS_MMAPPED(M)**: This bit is set when chunk is mmap'd.
  - **NON_MAIN_ARENA(N)**: This bit is set when this chunk belongs to a thread arena.

*B. Heap Memory Corruption Classification*

For heap memory corruption, it will involve two regions of memory, one source chunk and one or more target chunks. Target chunks can include application data or heap metadata.The attacker's goal is to overwrite some part of target chunk with attacker-controlled data. However, this is not only heap memory corruption. We do the deeper study about heap memory exploitation in the following.

**Inter-chunk heap overflow** involves two regions of memory, and it can be the most common vulnerability class, writing linearly off the end of a heap allocation. There was one very famous case off-by-one heap overflow in glibc [11]. This case can make heap end up in a deterministic state at the time of the exploit attempt. However, far more common is where the attacker is attacking a heap which is in a completely unknown state, such as in the context of a remote service, a web browser renderer process or a kernel.

**Intra-chunk heap overflow** [12] can provide a very powerful exploitation primitive. This memory corruption does not cross a heap chunk, and it just crosses the fields in the data structure. In this attack, all of the uncertainty and non-determinism arising from unknown heap state is eliminated. The heap can be in any state, yet the same program data will always be corrupted in the same way. Let us to see one code sample as follows. This simple code shows us that intra-chunk heap overflow is very powerful and reliable. And it is also very difficult to detect this memory corruption.

```
struct goaty {char name[8]; int should_run_calc;};
int main(int argc, const char* argv[]){
  struct goaty* g = malloc(sizeof(struct goaty));
  g->should_run_calc = 0;
  strcpy(g->name, "projectgoat");
  if (g->should_run_calc) execl("/bin/gnome-calculator", 0);
}
```

**Use-After-Free** is a very serious and reliable exploit. The reason is that their existence is a result of the combined actions from different parts of an application  namely, the parts of the code that can cause the freeing of the object and the parts of the code that use the object. Use-after-free heap memory corruption requires a thorough code review to be identified manually especially in complex and large codebases.

### III. SELFIE OVERVIEW

Sensitive memory is different with critical memory [13] which allows programmers to identify and protect data that is critical to the correct execution of their application. In some sense, the programmers can pre-define the sensitive memory. However, different users may have different requirements for sensitive memory. So it is better to define the sensitive memory by users.

Selfie can allow users to choose the sensitive memory in runtime. Whenever the users think about some content that should be sensitive, they can tag the memory that should be sensitive in the time window. When Selfie detects there is one heap memory corruption, the first thing will be to decide whether the memory corruption belongs to the sensitive memory. If it is sensitive memory, we must do execution recovery. If it is non-sensitive memory, we need firstly to figure out whether the memory corruption will affect the sensitive memory in future and then consider whether execution recovery is necessary.

Figure 1 presents the high-level overview of Selfie and details its five main components: (i) a modified memory allocator, (ii) dynamic instrumentation detection tool, (iii) a root cause analyzer, (iv) a symbolic solver and (v) execution recovery.

Selfie relies on a modified memory allocator to collect the sensitive memory. If Selfie does LD_PRELOAD for this modified memory allocator, it will replace the memory allocator in the beginning. So the whole of memory will be sensitive. So Selfie provides runtime libhijack feature which can make users to define which memory will be sensitive in runtime. Dynamic instrumentation tool will continue to check whether there are memory corruptions. Once memory corruptions happen, dynamic instrumentation tool will provide information whether it belongs to sensitive memory or non-sensitive memory. If it is non-sensitive memory, it doesn't mean we don't need consider this memory corruption. We need to figure out whether it can affect the sensitive memory by symbolic solver in future. If it does affect or it is sensitive memory, the root cause analyzer will figure out which data structure has been corrupted and which input make the memory corruption. Once Selfie figures out the root cause, Selfie can locate the clean state and then do execution recovery.

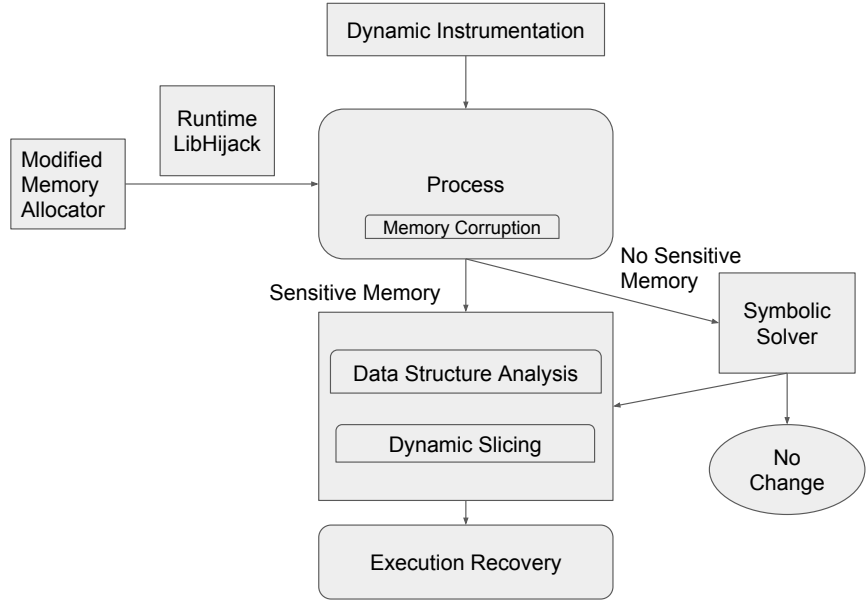

Fig. 1: Overview of Selfie

## IV. USER-DEFINED SENSITIVE MEMORY

Developers can define the critical memory based on the program logic. However, whether the memory is sensitive for users, and it can be different according to different users' requirements and different contexts. Let us think about one very simple case - text editor 'geany'. Geany provides many editor windows for different files at the same time. It is not practical for developers to make all editor windows sensitive. So the better solution is that the users can choose which editor window should be sensitive or include sensitive memory for users. Selfie provides one modified memory allocator to help users to define which memory should be sensitive in one time window and provide one tool to help user turn on/off the sensitive memory option.

### A. Modified Memory Allocator

Currently, Selfie focuses on glibc memory allocator(e.g. malloc, calloc and realloc). Selfie's goal is to avoid memory corruption for sensitive memory, and we don't consider whether the sensitive memory will leak. So we don't provide isolation or shadow memory for sensitive memory. We keep the sensitive memory integrity and we don't loose the sensitive memory once memory corruption.

Selfie overrides the glibc memory allocator and adds landmark to the end of the allocated memory. Dynamic instrumentation module can figure out whether the memory is sensitive memory based on the post landmark. Especially, symbolic solver needs to figure out whether there are illegal write to sensitive memory.

### B. Switch for Sensitive or Non-Sensitive

If users adopt the modified memory allocator in pre-load section, the whole heap memory will be sensitive and users cannot decide when to turn on/off the sensitive memory. So Selfie can provide the runtime libhijack to override the glibc memory allocator in runtime.

Selfie balances the libhijack to do good thing, change modified memory allocator in runtime. We really don't do code injection. Instead, Selfie allocates a new memory mapping and store auxiliary data in mapping,such as lib.so and name of the fuction to hijack. Libhijack will call dlopen and dlsym and replace GOT entry with entry found via dlsym. The key point is about how to reimplement dlopen. The high level will be load dependencies (deps can be loaded via real dlopen). And then it can create memory map and write .so data to new memory maps. Patch into the RTLD and run init routines can be the best way to figure out the address of GOT. Hijack GOT is the best way to override the memory allocator. Once the users want to disable the sensitive memory, Selfie just load the normal memory allocator.

## V. ROOT CAUSE FOR HEAP MEMORY CORRUPTION

It is always a challenge task to figure out the root cause for memory corruption. Especially, it is expensive to collect and store dynamic traces which may be needed to locate the memory corruption. Fine-grained tracing is needed so that the fault can be later analyzed and the root cause of the memory corruption can be detected. However, it is far more practical. Selfie integrates online dynamic and offline static analysis to figure out the root cause of memory corruption.

### A. Heap Memory Corruption Detection

There has been many solutions [14], [15], [16] about how to detect the heap memory corruption. Selfie adopts dynamic instrumentation tool to detect the heap memory corruption dynamically. Selfie will maintain two lists for memory allocation, one is sensitive memory allocation list and another one is non-sensitive memory allocation. Meanwhile, Selfie also maintains one free list. When application allocates an area in the heap, generally the memory allocator sets headers before the memory area (checking Section II-A). These headers can contain some information like size, used or not etc.. Basically, when Selfie monitor the allocator, it can determine the base address and the size of the allocation areas. When Selfie check all STORE and LOAD accesses in memory, it can determine which one is outside the two maintained memory allocation lists. For use-after-free, Selfie can easily check the free list to make sure whether the memory has been freed.

In Section II-B, we have mentioned two very interesting heap overflow: inter-chunk heap overflow and intra-chunk heap overflow. It is every easy to detect inter-chunk heap overflow based on the above solution. However, intra-chunk heap overflow is not easy to detect. The problem is that when Selfie records all memory allocation. The semantic information of memory has been discarded. Selfie just can maintain the base address and size of the memory allocation. But it cannot figure out the intra-chunk information. Semantic missing will directly miss the memory corruption. Even for [17], [18], they places canaries between chunks in the heap, but that too focuses on inter-chunk overflows rather than intra-chunk overflows. So intra-chunk heap overflow is still very serious problem. How to maintain the data structure semantic information in runtime will be still a promising direction. Based on the data structure information, field slicing (such as some data structure filed work [19]) may be one direction to figure out the intra-chunk heap overflow.

### B. Symbolic Solver

For the long running program, memory corruptions can introduce huge damage, but not all memory corruptions and some memory corruptions may be no any effect for the whole execution. The users only care about the correct execution and sensitive memory. If some memory corruptions don't affect the execution and sensitive memory, it is very expense to deal with them. Selfie use the concolic execution which is a technic that uses both symbolic and concrete execution to solve a constraint path. The concolic execution is mainly used to cover the effect of memory corruption in future.

### C. Corrupted Data Structure Extract

When Selfie detect the heap memory corruption, it will report which instruction make the memory corruption. We assume the application with debug information. So it will be easy to figure out the the line number corresponding with source code. How to automatically figure out the data structure name and definition according to line number? Selfie builds one clang plugin to automatically figure out the related data structure information. It can provide Selfie with the name of allocated data structures. To mark individual memory addresses with their fine-grained data types, Selfie needs to extract definitions of data structures from source codes. Selfie uses the Clang compiler front-end [20] to parse the source files and generate the corresponding abstract syntax trees. Selfie's plugins uses LibClang and LibTooling libraries to recursively traverse the generated syntax trees and extract data structure definitions automatically. Direct parsing of the individual source codes independently is not straightforward, because the sources contain several interdependencies such as header file inclusions as well as source-defined macros that need to be resolved before the parsing. Correct resolution of source file interdependencies is crucial for Selfie's accuracy, because we faced many different data structures with identical names defined in different unrelated files within the source code. Selfie employs Bear [21] to create a Clang JSON compilation database. Selfie extracts and captures all inter-file dependencies and resolves the macros. Consequently, Selfie uses the compilation database to parse correct source files for data structure definitions when the definitions do not reside in the same file where the structures are used.

### D. Dynamic Slicing

Data structure information let us know which data structure meets memory corruption. However, we still don't know which input or which step introduces the memory corruption. If we cannot figure out this problem, it will be very difficult to locate the clean state and do execution recovery. It is impossible to analysis each instructions from the beginning. It is better that we can just analyze the related instruction. Selfie uses program slice which contains all statements in a program that directly or indirectly affect the value of a variable or instruction. We can further narrow down the notion of the slice to dynamic program slice [22]. Unlike traditional slicing named static slicing which computes all statements that may have affected the value of a variable at a program point for any arbitrary execution of the program, the dynamic program slicing computes all statements that actually affect the value of a variable/instruction at a program point for a particular execution of the program with given input.

Selfie uses the dynamic program slicing to narrow down the analysis range. The lastest input instruction will be very possible reason for heap memory corruption. Once Selfie recognizes the root cause of memory corruption, Selfie can report where is the clean state, so it can make execution recovery from this clean state and avoid the bad input. Finally, Selfie benefits online dynamic and offline static analysis to do execution recovery not just exit and loose the sensitive memory.

## VI. EXECUTION RECOVERY

Memory checkpointing and recovery [23], [24], [25] has been very widely to improve the reliability of today's software stack. Because it is an important technique that allows users to snapshot the memory image of a running program in main memory and restore the checkpointed image later on. Most application scenarios require high checkpointing frequencies. For example, automatic error recovery techniques typically checkpoint the active memory image at every client request or at carefully selected rescue points, commonly resulting in thousands of checkpoints per second. Selfie takes snapshot based on function level. During the execution, it is possible for Selfie to take a snapshot of the registers and memory states in the entry of each function. Then, at each program point, it is possible to restore the previous snapshot. We don't means keep all the snapshot, because it will be very expense for memory. Though we do function level snapshot, we don't keep all the snapshots. We only focus on source code defined functions. Meanwhile, we analyze the function call graph and choose which function snapshot should be kept following the time window. Figure 3 shows one example about function call graph. We can see main function is the entry, and there are many different exit points. From each entry to exit, it is as one function call path. We keep each path differently. Based on the different paths, we choose whether some function snapshots should be kept.

## VII. IMPLEMENTATION

Selfie architecture, as presented in Figure 1, has been implemented on Linux v3.13 for x86_64 Ubuntu 14.04 machine without ASLR. Selfie integrates and modifies current research tools and develops new tools to finish final goal. Runtime libhijack is to balance libhijack [26] which has good support for FreeBSD but not for Ubuntu. I modify the code to make it work well on Ubuntu to load library by dlopen runtime. I modify glibc memory allocator (malloc, calloc and realloc) to add landmark information to override the normal memory allocator for sensitive memory.

Dynamic instrumentation tool is very useful for detecting heap memory corruption. In the beginning, I choose to use PIN [27] to do dynamic instrumentation. Later, I found out another tools Triton [28] which is a Pin-based dynamic symbolic execution (DSE) framework. Although Triton is a DSE framework, it also provides internal components like a taint engine, a snapshot engine, translation of x86 and x86-64 instructions into SMT2-LIB, a Z3 interface to solve constraints and, the last but not least, Python bindings. Based on Triton, Selfie build its own program analysis tool to finish memory corruption detection, symbolic solver and execution recovery. For recording sensitive memory tracing, we build the Read/Write memory tracer based on Triton. So we can know which instruction read or write the sensitive memory. However, runtime tracing based on PIN has super high performance overhead. As one option, I do offline analysis based on IDA Pro. Because we assume without ASLR (it is not good assumption). Selfie use one IDA Pro script to record all heap memory access across an executable using IDA debugger API. So Selfie can build the profile about heap memory access.

Root cause analyzer includes data structure analysis and related instructions analysis. It belongs offline analysis. In Section V-C, we have mentioned that Selfie develop one Clang Libtooling plugin to figure out data structure definition. Before data structure analysis, we firstly use gdb to figure out the line number of error instruction. About dynamic slicing, Selfie balances the Giri [29] which handles both data-flow and control-flow dependences when computing the dynamic backwards slice. Giri takes advantage of the LLVM intermediate representation(IR), which is static single assignment (SSA) form, to reduce the size of the trace file.

Our current Selfie prototype's detection and recovery is based on Triton dynamic symbolic framework. So it inherit the limitation of Triton.

## VIII. EVALUATION

We evaluated Selfie on Intel i7-4710MQ CPU 2.50GHz and 16 GB of RAM. We ran all our tests on an Ubuntu 14.04

Fig. 2: Selfie works well on simple off-by-one heap overflow

Linux kernel 3.13. We designed a set of experiments to verify whether Selfie can be useful in real-world practical scenarios. We firstly go through one simple code example and then target the real-world application.

### A. Case study: off-by-one heap overflow

Let us firstly focus on one very simple off-by-one heap overflow, and how Selfie detect the heap overflow and finish the execution recovery. In the following sample, it can have off-by-on heap overflow when n ¿= BUFF_MAX.

```
#define BUFF_MAX 128
int main(int ac, const char *av[])
{
  char *buff1, *buff2;
  int i;
  buff1 = (char*)malloc(BUFF_MAX);
  buff2 = (char*)malloc(BUFF_MAX);
  int n;
  scanf("%d", &n);
  if (!buff1 || !buff2)
    return -1;
  for (i = 0; i <= n; i++) /* off-by-one */
    buff1[i] = 'A';
  for (i = 0; i <= n; i++) /* off-by-one */
    buff2[i] = 'B';
  free(buff1);
  free(buff2);
  return 0;
}
```

Figure 2 shows that Selfie works well on this simple off-by-one heap overflow. The first time, we set the n value is 128. We can see there are heap overflow and Selfie recognizes the 128 is bad input. So Selfie requests the user to do second time input. When the user input 56 which is n's value. Now, there isn't any error. Selfie evaluates the input is good input. We can see that Selfie perfectly detects the heap overflow and then finishes execution recovery. There is many intermediate result about gdb recognizing line number, AST Tree Based on clang libtooling, data structure extracting and dynamic slicing, you can check code package.

### B. Case study: NullHttpd

Null httpd is a very small, simple and multithreaded web server for Linux and Windows, Which contains a remote exploitable heap overflow [30], when negative Content-Length values are transmitted. Next, we will show how Selfie detect the heap overflow. In this case study, we will analyze the heap overflow from memory aspect.

In the following, it shows us three pieces of memory dump. In the first piece, it is very easy for us to locate "ef ef ef ef fe fe fe fe" which is the landmark to tag the sensitive memory. In the third line, we can find hex sequences "10 67 42 b7" which is the size of calloc(2200). In Section II-A, we have illustrated the metadata about memory allocation. "10 67 42 b7" is the size of allocated memory, and next it will be the allocated memory. The second piece of memory dump show us the heap overflow location. We can check http.c file in line 100. If we just analyze the source code, the size of calloc should be more than 1024, however, the allocated size just 224. So if the buffer is larger than 224, there will be heap overflow. So the reason may locate in_ContentLength. Because in_ContentLength is the filed of conn[sid].dat (data structure analysis gives us more information about this data structure), If we find out the conn[sid].dat heap location, we can figure out the value of in_ContentLength. In third piece of memory dump, we find out the location of conn[sid].dat. And then we figure out the address of in_ContentLength which is 0xffffffce0, and the value is -800. So that means there is negative length. However, the negative length is not valid length. So negative length results in heap overflow.

```
sever.c
    462: conn=calloc(config.server_maxconn, sizeof(CONNECTION));
    calloc(2200) = 0x9792810
00000810  40 eb 75 b7 40 eb 75 b7  04 00 00 00 02 00 c1 22  |@.u.@.u........"|
00000820  7f 00 00 01 00 00 00 00  00 00 00 00 2b 47 46 56  |............+GFV|
00000830  2b 47 46 56 10 67 42 b7  68 04 40 b7 40 fb 57 b7  |+GFV.gB.h.@.@.W.|
00000840  40 fb 57 b7 05 00 00 00  02 00 c1 23 7f 00 00 01  |@.W........#....|
00000850  00 00 00 00 00 00 00 00  2f 47 46 56 2f 47 46 56  |......../GFV/GFV|
00000860  00 00 00 00 68 04 20 b7  00 00 00 00 00 00 00 00  |....h. .........|
00000870  00 00 00 00 00 00 00 00  00 00 00 00 00 00 00 00  |................|
*
000010a0  00 00 00 00 00 00 00 00  ef ef ef ef fe fe fe fe  |................|
000010b0  a4 08 00 00 00 00 00 00  00 00 00 00 00 00 00 00  |................|

heap overflow location:
http.c
    100: conn[sid].PostData=calloc(conn[sid].dat->in_ContentLength+1024, sizeof(char));
    calloc(224) = 0xb7426710

00026710  0a eb 0a 2d 2d 6e 65 74  72 69 63 2d 2d 31 c0 31  |...--netric--1.1|
00026720  db 31 c9 31 d2 b0 66 b3  01 51 b1 06 51 b1 01 51  |.1.1..f..Q..Q..Q|
00026730  b1 02 51 8d 0c 24 cd 80  b3 02 b1 02 31 c9 51 51  |..Q..$......1.QQ|
00026740  51 80 c1 77 66 51 b1 02  66 51 8d 0c 24 b2 10 52  |Q..wfQ..fQ..$..R|
00026750  51 50 8d 0c 24 89 c2 31  c0 b0 66 cd 80 b3 01 53  |QP..$..1..f....S|
00026760  52 8d 0c 24 31 c0 b0 66  80 c3 03 cd 80 31 c0 50  |R..$1..f.....1.P|
00026770  50 52 8d 0c 24 b3 05 b0  66 cd 80 89 c3 31 c9 31  |PR..$...f....1.1|
00026780  c0 b0 3f cd 80 41 31 c0  b0 3f cd 80 41 31 c0 b0  |..?..A1..?..A1..|
00026790  3f cd 80 31 db 53 68 6e  2f 73 68 68 2f 2f 62 69  |?..1.Shn/shh//bi|
000267a0  89 e3 8d 54 24 08 31 c9  51 53 8d 0c 24 31 c0 b0  |...T$.1.QS..$1..|
000267b0  0b cd 80 31 c0 b0 01 cd  80 bf ff b0 ef bf ff b0  |...1............|
000267c0  ef bf ff b0 ef bf ff b0  ef bf ff b0 ef bf ff b0  |................|
*
00026870  ef bf ff b0 ef bf ff b0  ff ff ff ff fc ff ff ff  |................|
00026880  ef 34 f3 04 08 d1 fb 04  08 0a 00 00 00 00 00 00  |.4..............|
00026890  00 00 00 00 00 00 00 00  00 00 00 00 00 00 00 00  |................|
*
00047000

the reason: negative length
server.c
    516    conn[sid].dat=calloc(1, sizeof(CONNDATA));
    calloc(78160) = 0xb7400468
    0xffffffce0 = -800

00000460  00 00 00 00 ad 62 02 00  00 00 00 00 00 00 00 00  |.....b..........|
00000470  00 00 00 00 00 00 00 00  e0 fc ff ff 00 00 00 00  |................|
00000480  00 00 00 00 00 00 00 00  00 00 00 00 00 00 00 00  |................|
*
000009f0  00 00 00 00 00 00 00 00  00 00 00 00 48 54 54 50  |............HTTP|
00000a00  2f 31 2e 30 00 00 00 00  00 00 00 00 00 00 00 00  |/1.0............|
00000a10  00 00 00 00 00 00 00 00  00 00 00 00 00 00 00 00  |................|
*
00000e80  00 00 00 00 00 00 00 00  00 00 00 00 31 32 37 2e  |............127.|
00000e90  30 2e 30 2e 31 00 00 00  00 00 00 00 00 00 00 00  |0.0.1...........|
00000ea0  50 4f 53 54 00 00 00 00  2f 00 00 00 00 00 00 00  |POST..../.......|
00000eb0  00 00 00 00 00 00 00 00  00 00 00 00 00 00 00 00  |................|
*
000013c0  00 00 00 00 00 00 00 00  00 00 00 00 ff ff ff ff  |................|
000013d0  00 00 00 00 00 00 00 00  00 00 00 00 00 00 00 00  |................|
*
000135b0  00 00 00 00 00 00 00 00  ef ef ef ef fe fe fe fe  |................|
000135c0  5c 31 01 00 00 00 00 00  00 00 00 00 00 00 00 00  |\1..............|
```

## IX. RELATED WORK

This section provides a general overview of existing memory protection and recovery techniques. We focus here on user-level memory protection and recovery.

### A. Checkpointing

Memory Checkpointing is a pivotal technique in systems reliability, with applications ranging from crash recovery to replay debugging. There are different checkpointing solutions like fork-based checkpointing [23], [24], [25], Mprotect-based checkpointing [31], [32], Soft Dirty Bit-based Checkpointing [33] and undolog-based checkpointing [34], [35], [13]. These memory checkpointing solutions require high-frequency checkpoints. So they incur either substantial runtime performance overhead, or poor memory usage guarantees. Though [13] proposes one lightweight memory checkpointing which combines low performance overhead with strong memory usage guarantees for high checkpointing frequencies. However, from users' aspect, they only care about their sensitive memory which defined by themselves. For example, users just want to protect and recover sensitive memory data in one important tab in Chrome or Firefox browser, and they don't care about other tab's memory. We provide one feasible solution for users who want to protect and recover self-defined sensitive memory. There is no doubt that our solution can reduce the runtime performance overhead and memory space.

### B. Error Tolerance

DieHard [10] is a memory allocator that hardens the application to memory errors. It can provide probabilistic soundess guarantees for applications that allow them to continue executioin in thee presence of memory errors. DieHard provides one replicated mode, which means the whole process is replicated. The replicated process still will incur performance overhead and is not feasible for users. Samurai [36] adopts critical memory concept and only replicates critical memory. Though it can provide strong protection guarantees for the application, it needs programmer effort and the critical memory generally is not sensitive memory for users.

### C. Memory Tracing and Watchpoints

Memory tracing allows the execution of memlets for each memory access. MemTrace [37] provides dynamic lightweight memory tracing for unmodified binary applications. MemTrace adds shadow memory and state for each memory address of an application and allows the execution of user-defined memlets to inspect memory accesses. The serious problem is that MemTrace [37] uses cross-ISA translation for 32-bit applications to a 64-bit ISA to offer both a wider address space and additional register to user-defined memlets. So MemTrace cannot work for 64-bit applications. Metric [38] is a memory tracing framework that collects and stores selected memory access traces. Metric extracts access traces based on dynamic binary rewriting of the execution application. [39] proposes one general framework hwo to analyze and trace memory and understand the memory layout. [40] provides one hardware-based memory meta data protection solution. Our solution, Selfie, just need dynamic load or unload our memory allocation lib for user-defined sensitive memory.

*D. Taint Tracking and Data Flow analysis*

Taint tracking and data flow analysis extend memory tracing and analysis the flow of data inside an application. Every memory cell and every register has an associated tag. TaintCheck [41] propose dynamic taint analysis for automatic detection of overwrite attacks, which include most types of exploits. TaintCheck doesn't need source code or special compilation for the monitored program. DataFlowSanitizer [42] is a generalised dynamic data flow analysis. Though TaintCheck doesn't need source code, it incur high runtime overhead. DataFlowSanitizer need source code and compiler extension to add tag information in the code. They track the whole memory while the users may just interest in sensitive memory.

*E. Binary Translation*

Binary translation is to instrument a binary application. For example, IR-based binary translators translate the application by using a traditional compiler approach. It can add the desired instrumentation and generates machine code for the desired instrumentation, such as DynamoRIO [43], PIN [27], QEMU [44], and Valgrind [9]. Instrumentation will incur more performance overhead. Meanwhile, it is no easy for user to choose user-defined sensitive memory. Selfie can provide much more feasible for users.

*F. Memory Analysis and Recovery*

Reviver [45] proposes one memory data structure instance recovery framework to analyze the memory layout. [46] makes one memory data attack once learning about the memory layout. Dscrete [47] provides memory rendering through reuse of application logic to reverse engineer a memory image for more semantic information. Dscrete heavily relies on exact identification of the so-called P function within the binary that takes as input the target data structure instance and produce the human readable application output. Dscrete tries to figure out which memory is useful and to recover. However, Selfie can provide the user-defined sensitive memory and narrow the range of recovery.

## X. CONCLUSION

Different users always have different requirement for sensitive memory definition. It is not flexible for aborting program execution once detecting memory corruption. Because the users may loose some sensitive data. We presented Selfie, a hybrid solution to provide one flexible solution to protect the sensitive memory according to users' requirements in runtime. In addition, Selfie provides one comprehensive solution to detect the heap memory corruptions and figure out the root cause of memory corruption by data structure extracting and dynamic program slicing. Further, Selfie will decide whether the execution recovery is necessary.

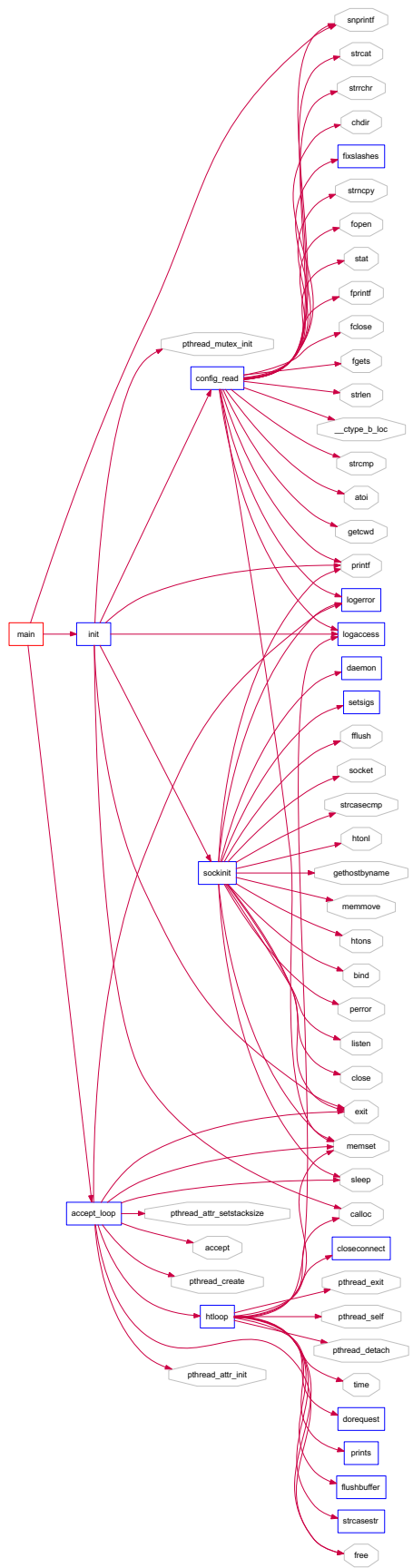

Fig. 3: Function Call Graph for NullHTTPD